\documentclass[12pt]{article}
\usepackage{amsfonts,amsthm,amsmath,amssymb,upgreek,bm}
\usepackage{hyperref}
\usepackage[paper=letterpaper,margin=.85in]{geometry}
\usepackage{graphicx}
\usepackage{units}
\usepackage{float}
\usepackage[export]{adjustbox}
\usepackage{bbold}
\usepackage{xcolor}
\usepackage[shortlabels]{enumitem}
\usepackage{dsfont}

\newcommand{\be}{\begin{equation}}
\newcommand{\ee}{\end{equation}}
\renewcommand{\tilde}{\widetilde}
\renewcommand{\i}{\mathrm{i}}
\renewcommand{\d}{\mathrm{d}}

\newcommand{\Ai}{\text{Ai}}
\newcommand{\Bi}{\text{Bi}}
\newcommand{\SLtwo}{\text{SL}(2,\mathbb{R})}
\newcommand{\density}{\uprho}

\numberwithin{equation}{section}

\begin{document}
\thispagestyle{empty}

\vspace*{.5cm}
\begin{center}

{\bf {\LARGE Finite-cutoff JT gravity and self-avoiding loops}}

\begin{center}

\vspace{1cm}

 {\bf Douglas Stanford and Zhenbin Yang}\\
  \bigskip \rm
  
\bigskip
Stanford Institute for Theoretical Physics,\\Stanford University, Stanford, CA 94305

\rm
  \end{center}

\vspace{2.5cm}
{\bf Abstract}
\end{center}
\begin{quotation}
\noindent

We study quantum JT gravity at finite cutoff using a mapping to the statistical mechanics of a self-avoiding loop in hyperbolic space, with positive pressure and fixed length. The semiclassical limit (small $G_N$) corresponds to large pressure, and we solve the problem in that limit in three overlapping regimes that apply for different loop sizes. For intermediate loop sizes, a semiclassical effective description is valid, but for very large or very small loops, fluctuations dominate. For large loops, this quantum regime is controlled by the Schwarzian theory. For small loops, the effective description fails altogether, but the problem is controlled using a conjecture from the theory of self-avoiding walks.

\end{quotation}

\setcounter{page}{0}
\setcounter{tocdepth}{2}
\setcounter{footnote}{0}
\newpage

\tableofcontents

\pagebreak

\section{Introduction}
In Jackiw-Teitelboim (JT) gravity \cite{Teitelboim:1983ux,Jackiw:1984je,almheiri2015models}, the path integral on the disk topology reduces to an integral over simple closed curves $\gamma$ in hyperbolic space, weighted by the area enclosed:
\be\label{first}
Z(\upbeta) \sim \int\mathcal{D}\gamma \  e^{p \cdot \text{area}(\gamma)}, \hspace{20pt} \text{length}(\gamma) = \upbeta.
\ee
Typically, this problem is studied in a simplifying limit, where $\upbeta$ and $p$ go to infinity, with a fixed ratio. We will study the more general case with $\upbeta$ and $p$ finite. The goal of doing this is to explore some of the issues that can arise in defining quantum gravity with finite boundary conditions.\footnote{This ``finite cutoff'' problem is important in AdS/CFT, see  \cite{Heemskerk:2010hk} and \cite{Susskind:1998dq,deBoer:1999tgo,Faulkner:2010jy,Douglas:2010rc}. Recent proposals based on the $T\bar{T}$ deformation \cite{Smirnov:2016lqw} have been discussed following work by McGough, Mezei, and Verlinde \cite{McGough:2016lol,Kraus:2018xrn,Donnelly:2018bef,Hartman:2018tkw,Gorbenko:2018oov,Guica:2019nzm}.} This problem was previously studied in \cite{Kitaev:2018wpr,Yang:2018gdb}; the thing we will add is that we will analyze this integral including the constraint that $\gamma$ should not self-intersect.

Defining the length of the curve $\gamma$ is subtle. In the path integral, typical curves will have a fractal microscopic structure, with dimension $4/3$, and a curve of macroscopic size will have infinite length. In most of the paper, we will follow \cite{Kitaev:2018wpr} and set $\upbeta$ equal to a multiplicatively renormalized version of this microscopic length. Concretely, we introduce a UV cutoff and then define $\upbeta$ as a rescaled version of the length as measured at the UV scale. The rescaling is chosen in a local way, meaning that it cannot depend on the overall length of the curve or the pressure. Informally, this definition should be the same as defining $\upbeta$ as the $4/3$-dimensional Hausdorff measure of $\gamma$, although we will not attempt to use this perspective.

\begin{figure}
\begin{center}
\includegraphics[width = .8\textwidth]{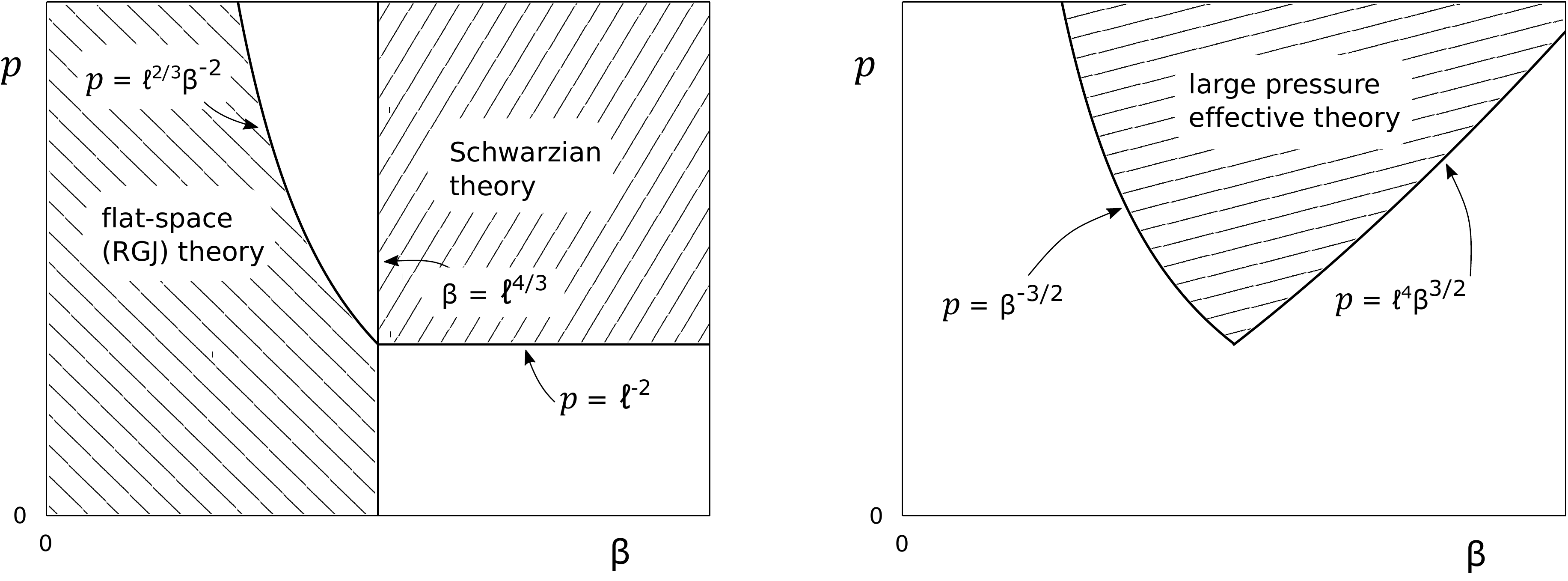}
\caption{{\small The three approximations used in this paper are valid well inside the respective shaded regions. The horizontal axis is the renormalized length of the loop, and the vertical axis is the pressure (or equivalently the boundary value of the JT gravity dilaton).}}\label{fig:regimes}
\end{center}
\end{figure}

We did not find an exact answer for the path integral (\ref{first}), but if the pressure is large in units of the curvature of hyperbolic space $p\ell \gg 1$ (this corresponds to a semiclassical, small $G_N$ limit in JT gravity), we will be able to accurately compute $Z(\upbeta)$ for any value of $\upbeta$, using three overlapping regimes. The regions of applicability of these regimes are sketched in figure \ref{fig:regimes}. 

In order to write the answers we got, it is convenient to think of $Z(\upbeta)$ as a thermal partition function, and to discuss the corresponding density of states, $\density(E)$, defined so that 
\be
Z(\upbeta) = \int \mathrm{d}E\,\density(E)e^{-\upbeta E}.
\ee
The three regimes apply in different ranges of the energy $E$. We will briefly describe them.
\begin{enumerate}
\item {\bf The Schwarzian regime} applies when the self-avoiding loop is much larger than the AdS scale. This corresponds to the regime of low energies, and the density of states is approximately
\be
\density(E) \approx \frac{\lambda^2}{2\pi^3}\sqrt{|E_0|}\sinh\left[\sqrt{6}\pi p\ell^2\sqrt{\frac{E-E_0}{|E_0|}}\right],\hspace{20pt} E_0 \le E \ll E_0 + p^{2/3}.
\ee
Here we have chosen the ground state $E_0$ to be at a negative value, proportional to $-p^{4/3}$ in AdS units. The constant $\lambda$ is arbitrary and will be discussed below, and $\ell$ is the curvature radius of the hyperbolic space.
\item {\bf An intermediate regime} applies for energies that are between $E_0$ and zero, but not too close to either end. In this regime, we use a large-pressure effective theory to describe the self-avoiding walk, with the microscopic wiggles in the shape of the boundary described by a nonlinear ``entropic tension'' term. By doing a one-loop computation in this theory, we find the density of states 
\be\label{densInt}
\density(E) \approx 
\frac{\lambda^2}{4\pi^3}\sqrt{|E|}\exp\left[2\pi p \ell^2\sqrt{1 - \Big(\frac{E}{E_0}\Big)^{3/2}}\right], \hspace{20pt} E_0 + p^{-2/3}\ell^{-8/3} \ll E \ll -p^{2/3}.
\ee
\item {\bf The flat-space regime} applies for energies that are either close to zero or positive. In this regime, the self-avoiding loop is very small compared to the curvature scale $\ell$. The problem can then be solved exactly thanks to a remarkable conjecture of Richard, Guttmann, and Jensen \cite{richard2001scaling}. The fact that we have an exact solution in this regime is important, because the loop becomes small enough that the large-pressure approximation breaks down, and no semiclassical treatment is possible. The exact answer leads to the following formula
\be\label{airyDensity}
\density(E) \approx \frac{\lambda^2}{4\pi^4}\frac{p^{1/3}e^{2\pi p\ell^2}}{\Ai^2(-E/p^{2/3}) + \Bi^2(-E/p^{2/3})}, \hspace{20pt} -p \ell^{2/3} \ll E < \infty.
\ee
\end{enumerate}
It is straightforward to check that these three formulas can be smoothly matched together in their overlapping regimes of validity. Together, they provide a consistent picture with a reasonable-looking and positive density of states. However, in the intermediate regime, the answer does not agree as might be expected with the partition function of classical JT gravity at finite cutoff. We will return to this point in the Discussion.

In the rest of the paper, we will review JT gravity, discuss the self-avoiding loop measure, and then analyze these three regimes in reverse order.

{\bf Note:} JT gravity at finite cutoff has also been studied by Iliesiu, Kruthoff, Turiaci, and H.~Verlinde using different methods \cite{TTBAR}. We are coordinating the submission of our papers.

\section{Brief review of JT gravity}
The basic variables of JT gravity are the metric $g_{\mu \nu}$ and a scalar field called the dilaton $\phi$. The action on a manifold $M$ is given by
\be
I_{JT}=-S_0\chi(M)-{1\over 2}\left[\int_M \phi \,(R+{2\over \ell^2})+2\int_{\partial M} \phi K\right]
\ee
where $\chi(M)$ is the Euler characteristic of $M$ and $\ell$ is the AdS radius. The coupling $S_0$ determines the weighting of different topologies. We will imagine that $S_0$ is very large, so that we only need to consider the leading topology. However, to declutter formulas we will actually omit the $S_0$ parameter below. It can be restored by multiplying $Z(\upbeta)$ and $\density(E)$ by $e^{S_0}$.

In AdS/CFT applications of JT gravity \cite{almheiri2015models,Jensen:2016pah,maldacena2016conformal, almheiri2016conformal,engelsoy2016investigation}, the most important thing to compute is the ``disk'' path integral. This is the path integral over topological disks with fixed boundary length and fixed boundary value of the dilaton:
\be\label{pathInt}
Z(L) = \int_{\substack{\text{length}(\partial M) = L \\ \phi|_{\partial M} = p\ell^2}} \mathcal{D}g_{\mu\nu}\mathcal{D}\phi\, e^{-I_{JT}}.
\ee
We are choosing the parametrize the fixed boundary value of the dilaton as $\phi = p\ell^2$. The parameter $p$ will be referred to as the pressure, for reasons that will become clear below.

\begin{figure}
\center
\includegraphics[width=.7\textwidth]{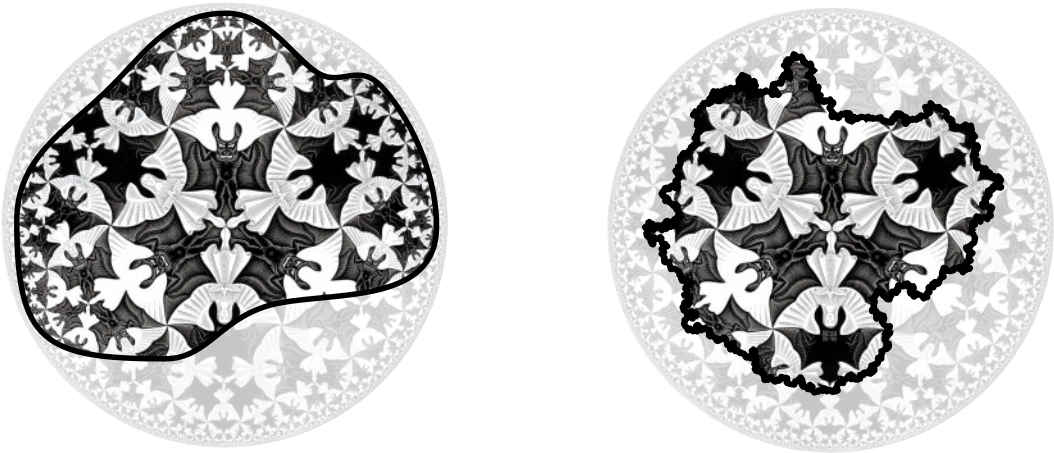}
\label{fig:cutout}
\caption{{\small Cutouts from the hyperbolic disk, bounded by a simple closed curve $\gamma$. Typical curves will be very wiggly at short distance scales, as shown at right.}}
\end{figure}

Since the bulk JT action is linear in the dilaton field, integrating out $\phi$ (along a contour parallel to the imaginary axis) imposes a delta function constraint on the metric $R+{2\over \ell^2}=0$. This means that the manifold $M$ is uniformly negatively curved.
Together with the restriction to the leading (disk-like) topology, this forces $M$ to be the interior of some simple closed curve $\gamma$ in the hyperbolic disk, see figure \ref{fig:cutout}.

On such a configuration, the JT action is reduces to the Gibbons-Hawking term, which can be simplified using the Gauss-Bonnet theorem:
\be
I_{JT}=-p\ell^2\int_{\partial M}K=-2\pi p\ell^2 \chi(M)+{p\ell^2\over 2}\int_M R=-2\pi p\ell^2-p A.
\ee
Since the curvature is a constant, its bulk integral gives us the area, $A$, of the region $M$. So the path integral (\ref{pathInt}) reduces to an integral over simple closed curves $\gamma$, weighted by the area enclosed:
\be\label{pathInt2}
Z(L) = \int_{\text{length}(\gamma) = L} \frac{\mathcal{D}\gamma}{\SLtwo}\, e^{p A + 2\pi p\ell^2}.
\ee
In this integral, we quotient by the action of $\SLtwo$ on the hyperbolic disk, which moves $\gamma$ around but does not change its geometry. In other words, by parameterizing $M$ in terms of a shape that is cut out of the hyperbolic disk, we are introducing an $\SLtwo$ gauge redundancy, and in (\ref{pathInt2}) we are correcting for this.

Motivated by JT gravity, Kitaev and Suh \cite{Kitaev:2018wpr} and Yang \cite{Yang:2018gdb} analyzed a variant of this integral. To define the measure $\mathcal{D}\gamma$, they took the curves $\gamma$ to be a continuum limit of a random-walk problem, where $\gamma$ is built out of $N$ straight segments of length $a$. In the continuum limit, $a$ goes to zero and $N$ goes to infinity, and a renormalized version of the length is defined as
\be\label{betaORW}
\beta = \frac{N a^2}{2\ell},
\ee
which is held fixed. With this definition of the integral, the authors of \cite{Kitaev:2018wpr,Yang:2018gdb} showed how to compute (\ref{pathInt2}) by solving a diffusion equation for a particle in hyperbolic space with an imaginary magnetic field. The result of their computation was
\be \label{formula:Random Walk partition function}
Z(\beta,p)=\int_0^{\infty} \mathrm{d} E \,\density(E) e^{-\beta E}, \hspace{20pt} \density(E) = {1\over (2\pi)^2}{\sinh(2\pi \sqrt{2E\ell})\over \cosh(2\pi p\ell^2)+\cosh (2\pi \sqrt{2E\ell})} e^{2\pi p\ell^2}.
\ee

However, as noted in the original papers \cite{Kitaev:2018wpr,Yang:2018gdb}, this solvable random walk model is not quite the correct quantum definition of JT gravity. The reason is that $\gamma$ is supposed to be the boundary of a disk-like domain $M$. This means that it should not self-intersect. This constraint is not respected by the sum over ordinary random walks that can be analyzed using the heat equation. In the rest of the paper, we will try to define and analyze the integral (\ref{pathInt2}) including the self-avoiding constraint.

\section{JT gravity and the self-avoiding loop measure}
\subsection{Flat space JT gravity}
A version of JT gravity can be defined that gives flat space (rather than hyperbolic space) as a solution. After a field redefinition, this theory is equivalent to the CGHS model without matter fields, see appendix \ref{app: CGHS}. The path integral reduces to an integral over non-self-intersecting closed curves in flat space $\gamma$, weighted by their enclosed area:
\be\label{flatpathint}
Z(\beta) = \int_{\text{length}(\gamma)=\beta}\frac{\mathcal{D}\gamma}{\text{rot.$\times$trans.}} \ e^{p\,A}.
\ee
As a warmup for the case of ordinary (hyperbolic) JT gravity, we would like to understand what measure to use to integrate over $\gamma$.
\begin{figure}
\begin{center}
\includegraphics[width = .5\textwidth]{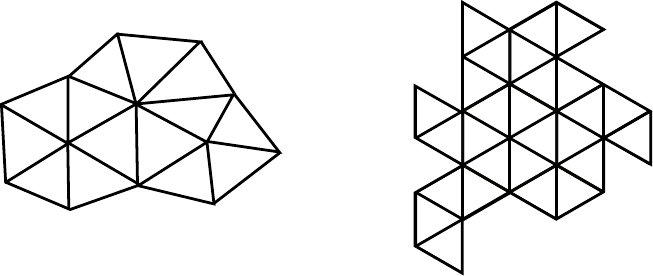}
\caption{{\small Random triangulations give a discrete regularization of the path integral over 2d metrics. All triangles are taken to be equilateral, and curvature is localized at the vertices. Positive curvature means fewer than six edges meeting at a vertex, and negative curvature means more than six. A discretization of flat-space JT gravity could be defined as a sum over triangulations in which each vertex meets exactly six edges. This reduces to a sum over boundary shapes on the triangular lattice (right).}}\label{fig:triangulations}
\end{center}
\end{figure}

A successful approach to defining quantum gravity in two dimensions is to define a discrete version of a path integral over metrics by the sum over all possible triangulations \cite{Ambjorn:1985az,David:1984tx,Kazakov:1985ea}. In the simplest version, all triangles are taken to be equilateral, so that the curvature is concentrated at the vertices
\be
\int \sqrt{g}R \longrightarrow \sum_{\text{vertices}}4\pi\left(1 - \frac{\text{\# edges meeting at vertex}}{6}\right).
\ee
The sum over geometries is then replaced by a discrete sum over triangulated graphs, and the measure $\mathcal{D}g_{\mu\nu}$ is replaced by a uniform weighting of all triangulations. In a continuum limit, this is believed to give the same measure as the one that follows from the ultralocal (DeWitt) metric, and the agreement between double-scaled matrix integrals and Liouville quantum gravity is evidence in favor of this.

Let's try to apply this approach to flat space JT gravity. The dilaton constrains the metric to be flat, and we can accomplish this by taking the number of edges at each vertex to be six. The boundary of the resulting triangulation is a non-self-intersecting loop on the regular triangular lattice, see figure \ref{fig:triangulations}. The sum over triangulations reduces to a uniform sum over such loops. The uniform measure on non-self-intersecting lattice walks or loops is a famous problem known as the ``self-avoiding random walk,'' and the argument above suggests that we should define $\mathcal{D}\gamma$ as the measure associated to the continuum limit of the self-avoiding walk. 

\begin{figure}
\begin{center}
\includegraphics[width = .65\textwidth]{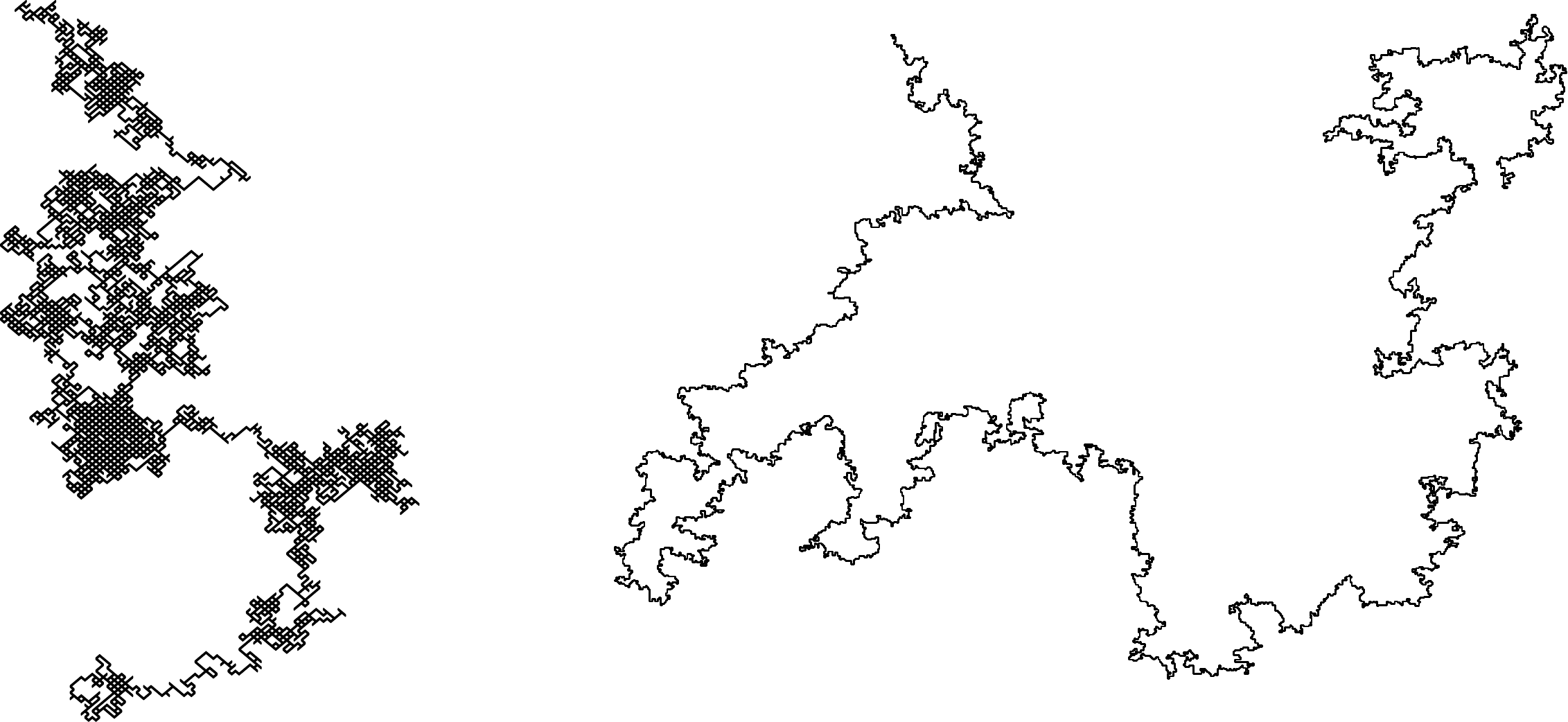}
\caption{{\small An ordinary random walk (left) and a self-avoiding walk (right) of $10^4$ steps each.}}\label{fig:walks}
\end{center}
\end{figure}

Microscopically, a typical self-avoiding walk is a fractal, and the total length is infinite. This means that as we take the continuum limit, we will have to hold fixed some kind of rescaled length, as for the case of the ordinary random walk (\ref{betaORW}). A clue for what to hold fixed is the following. For a typical self-avoiding random walk, linear measures of the size scale with the number of steps and the step size as
\be\label{sizes}
\text{size} \sim a\, N^{3/4}, \hspace{20pt} a = \text{lattice spacing},\hspace{20pt} N = \text{\# steps},
\ee
See figure \ref{fig:walks} for example configurations. In practice, the coefficient of proportionality in (\ref{sizes}) is non-universal and depends on the choice of lattice. It is convenient to absorb this dependence into a parameter $c_1$. To be concrete, we will choose $c_1$ so that the expected value of the area enclosed by a loop is
\be
\langle A \rangle = \frac{\sqrt{\pi}}{2}\, a^2\left(\frac{N}{c_1}\right)^{3/2}.\label{c1def}
\ee
The numerical coefficient was chosen to simplify some formulas below.

In order to take a continuum limit, one would like to take $N$ to infinity and scale the lattice spacing $a$ to zero in such a way that simple ``size'' quantities remain finite. To ensure this, we can hold fixed the combination 
\be\label{defupbeta}
\upbeta = a^{4/3}\frac{N}{c_1}.
\ee
This quantity is proportional to the number of steps, and we will take it to define the renormalized UV length of the curve $\gamma$.\footnote{An unusual feature is that this ``length'' has dimensions of length$^{4/3}$. In a similar situation, \cite{Kitaev:2018wpr,Yang:2018gdb} restored dimensions using the AdS length. We will just work with an $\upbeta$ with units length$^{4/3}$, and leave it to the reader to insert a dimensional factor if they want energies and lengths to have familiar units.}
With this notion of length fixed, one can now define the partition function of JT gravity as
\begin{align}\label{proposal}
Z(\upbeta) &= \lim_{N\to\infty} \mathcal{N}_N\hspace{-15pt}\sum_{\substack{\text{self-avoiding}\\\text{loops of }N\text{ steps}}}\hspace{-10pt}e^{p\, A}.
\end{align}
In this formula, the sum is over {\it rooted} self-avoiding loops, in which we choose a marked point on the loop that starts at the origin of the lattice. The normalization factor is
\be\label{factors}
\mathcal{N}_N = e^{-\# N}\frac{1}{2\pi}\frac{\lambda^2}{a_0}.
\ee
The constant $\#$ is an order-one number, which must be tuned so that the limit is finite and nonzero. The remaining factors are related to implementing the gauge-fixing of rotations and translations in (\ref{flatpathint}). In the continuum we could account for the gauge redundancy by inserting in the path integral the factor
\be\label{contfact}
\frac{1}{2\pi}\lambda^2\delta(x)\delta(y)
\ee
where $x,y$ are the coordinates of the marked point on the loop. Here, $\lambda$ is an arbitrary parameter with dimensions of length, and in studies of ordinary JT gravity it is often taken to be $\ell$, the AdS length. The factors in (\ref{factors}) are a lattice version of this insertion, since if $a_0$ is the area per lattice site, then $(1/a_0)\delta_{x,0}\delta_{y,0}$ is a lattice version of $\delta(x)\delta(y)$.\footnote{Depending on the lattice, it might be that there are significantly different numbers of self-avoiding loops of even vs.~odd length. We define the limit in (\ref{proposal}) to average over this.}

\subsection{Ordinary (hyperbolic) JT gravity}
In standard JT gravity, the partition function in the disk topology reduces again to an integral over non-intersecting pressurized closed loops, this time in hyperbolic space. We expect that the measure can again be defined by a continuum limit of walks on some discrete approximation of hyperbolic space, so that the partition function is given by\begin{align}\label{proposal2}
Z(\upbeta) &= \lim_{N\to\infty} \mathcal{N}_N\hspace{-15pt}\sum_{\substack{\text{self-avoiding}\\\text{loops of }N\text{ steps}}}\hspace{-10pt}e^{p(A + 2\pi\ell^2)}, \hspace{20pt} \mathcal{N}_N = e^{-\# N}\frac{1}{2\pi}\frac{\lambda^2}{a_0}.
\end{align}
Relative to the flat space case, we have inserted an extra constant $2\pi p\ell^2$ in the exponent. This is to agree with (\ref{pathInt2}).

The sum is now over rooted self-avoiding loops on some (irregular) discrete approximation of hyperbolic space with radius of curvature $\ell$, and with typical lattice spacing $a$. Although for an appropriate choice of $c_1$, we expect the continuum limit to be independent of details of the lattice, for concreteness one can imagine that on scales much smaller than $\ell$, the discrete graph resembles a flat-space lattice with a known $c_1$. We then define $\upbeta$ as in (\ref{defupbeta}).

\subsection{The small \texorpdfstring{$\upbeta$}{beta} limit}
Let's test out these definitions in the limit $\upbeta \rightarrow 0$, where the loops will be small enough that the pressure term and the curvature are unimportant. Then $Z(\upbeta)$ is simply proportional to the number of $N$-step self-avoiding loops in flat space. This is believed to behave as
\be\label{refinement}
\sum_{\text{SAL}_N}1 \ = \frac{B}{N^{3/2}}e^{\# N},
\ee
where $B,\#$ are non-universal constants. Inserting this into (\ref{proposal}), one finds that for $\upbeta \rightarrow 0$, 
\be\label{betaToZero}
Z(\upbeta) = \frac{a^2B}{2\pi a_0 c_1^{3/2}}\frac{\lambda^2}{\upbeta^{3/2}}.
\ee
The prefactor in this expression is a combination of non-universal constants. This looks bad, because we would like the partition function on the LHS to be universal and independent of the details of the lattice. In fact, we are saved by the fact that this particular combination of factors has a universal answer \cite{cardy1994mean}, and in fact
\be
Z(\upbeta) = \frac{1}{8\pi^{5/2}}\frac{\lambda^2}{\upbeta^{3/2}}.
\ee
This gives the answer for the partition function of JT gravity in the large $\upbeta$ limit, either in flat space or hyperbolic space.\footnote{To be consistent with our conventions, in hyperbolic space we should multiply the answer by $e^{2\pi p \ell^2}$.}

In order to go beyond this limit of small $\upbeta$, we need a refinement of (\ref{refinement}) that includes the pressure and/or curvature. Remarkably, Richard, Guttmann and Jensen \cite{richard2001scaling} made a conjecture that solves precisely this problem for the case of flat space JT gravity at arbitrary pressure. We will discuss this next.

\section{The flat space regime}
One can imagine expanding the partition function of JT gravity in flat space
\be
Z(\upbeta) = \lim_{N\to\infty} \mathcal{N}_N\sum_{\text{SAL}_N}e^{p\, A}
\ee
in powers of the pressure $p$. The coefficient of $p^n$ will be proportional to $\langle A^n\rangle$, where the expectation value is taken at zero pressure. The expectation value $\langle A\rangle$ scales as $\upbeta^{3/2}$, see (\ref{c1def}), and one expects $\langle A^n\rangle$ to scale as $\upbeta^{3n/2}$. So the expansion in powers of $p$ will really be an expansion in powers of $\upbeta^{3/2}p$, and $Z(\upbeta)$ can be written as
\be\label{rhrrjrhrj}
Z(\upbeta) = \frac{\lambda^2}{8\pi^{5/2}}\frac{1}{\upbeta^{3/2}}\,f(y), \hspace{20pt} y = \upbeta^{3/2}p,
\ee
for some function $f$. 

In \cite{richard2001scaling}, Richard, Guttmann and Jensen (RGJ) made a conjecture\footnote{By comparing to extrapolated exact enumeration data for the moments of the area of an unpressurized self-avoiding loops, the first ten terms in the expansion of $f(y)$  have been checked to roughly eight digits of precision each \cite{jensen2004enumeration}. Also, a derivation was suggested in \cite{cardy2001exact} based on the connection to branched polymer physics in the region of large negative pressure. Finally, we will give an independent (although much less precise) check of the formula in the large $y$ region below.} that implies that
\be\label{f}
f(y) = 2\sqrt{\pi}\sum_{n = 0}^\infty \frac{f_n}{\Gamma(\frac{3n-1}{2})}y^n,
\ee
where
\be\label{fn}
f_0 = -1, \hspace{20pt} f_n = \frac{3n-4}{4}f_{n-1} + \frac{1}{2}\sum_{k = 1}^{n-1}f_kf_{n-k}.
\ee
These $f_n$ coefficients appear in the large-argument expansion of the function $\Ai'(s)/\Ai(s)$. For large $n$, they grow factorially, proportional to $\Gamma(n)$. However, because of the $\Gamma$ function in the denominator of (\ref{f}), the series converges for any value of $y$. The small $y$ behavior follows directly from the definition, and the large $y$ asymptotics are discussed in appendix \ref{app:asymp}:
\be
f(y) = \begin{cases} 1 + \frac{\sqrt{\pi}}{2}y + \frac{5}{12}y^2 + \dots  & y \ll 1 \\ e^{y^2/12}\left[y^2 + O(1)\right] & y \gg 1.\end{cases}
\ee
\begin{figure}
\begin{center}
\includegraphics[width = .35\textwidth]{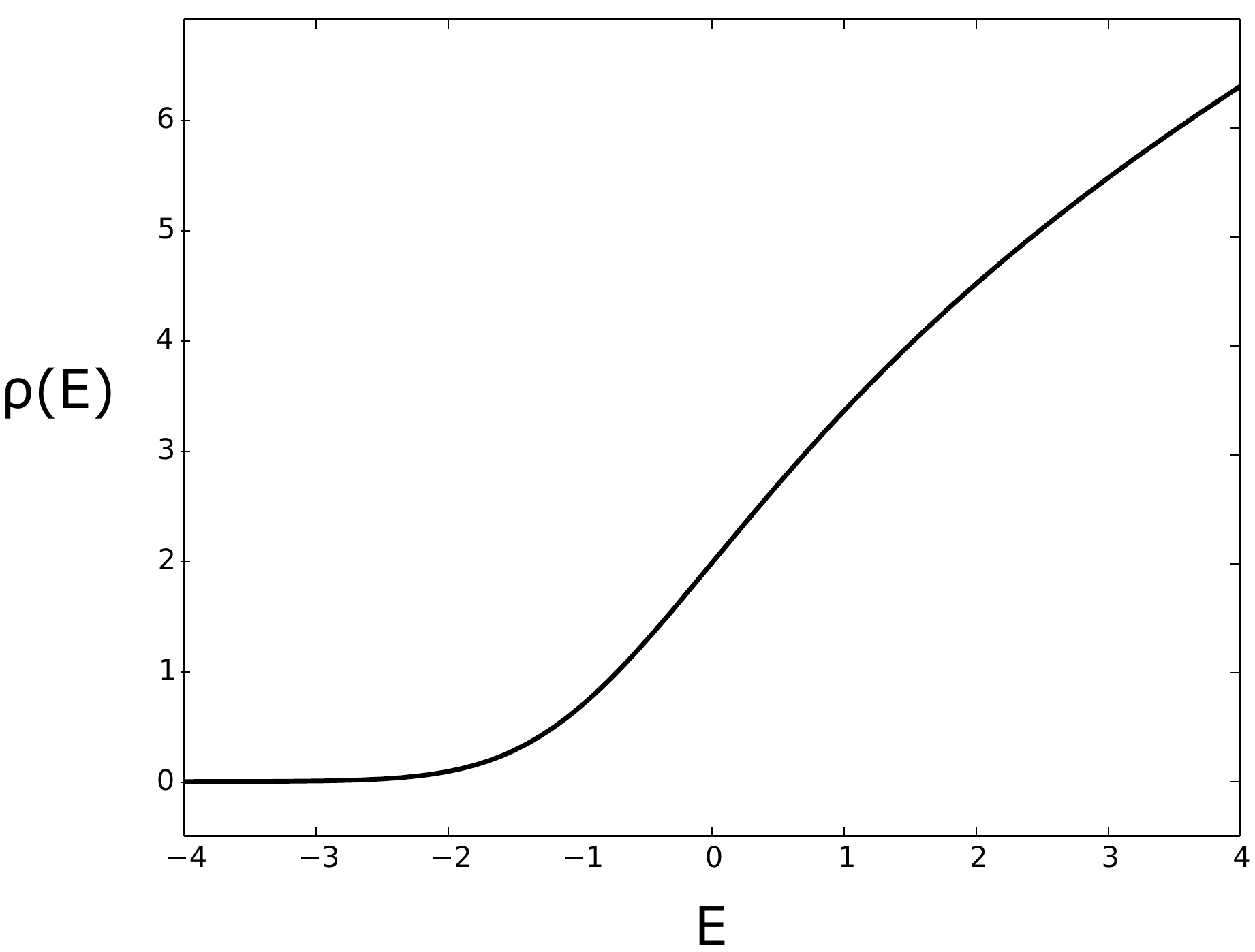}
\caption{{\small A plot of the function $[\Ai(-E)^2 + \Bi(-E)^2]^{-1}$. The region shown connects the behavior $\sqrt{E}$ for large positive $E$ to the behavior $e^{-\frac{4}{3}|E|^{3/2}}$ for large negative $E$.}}\label{fig:airy}
\end{center}
\end{figure}

This expression for $f(y)$ gives an exact solution to flat space JT gravity at finite cutoff, or equivalently the CGHS model without matter fields at finite cutoff. We would like to interpret the resulting formula for $Z(\upbeta)$ as a thermal partition function, and it is natural to ask what the corresponding density of states is. In appendix \ref{app:RGJ}, we show that
\be\label{rhoFlat}
Z(\upbeta) = \int_{-\infty}^\infty \mathrm{d}E \,\density(E) e^{-\upbeta E}, \hspace{20pt} \density(E) = \frac{\lambda^2}{4\pi^4} \frac{p^{1/3}}{\Ai^2(-Ep^{-2/3}) + \Bi^2(-Ep^{-2/3})}.
\ee
This spectrum is supported on the entire real axis. For large positive $E$, it grows as $\sqrt{E}$. This could have been anticipated based on the small $\upbeta$ behavior (\ref{betaToZero}), which represents the pressureless case. For large negative $E$, the density has a small tail that decays exponentially:
\be\label{ent1}
S(E) = \log\,\density(E) = -\frac{4}{3p}|E|^{3/2}, \hspace{20pt} (-E \gg p^{2/3}).
\ee
The answer in this limit can be obtained from a semiclassical argument based on a limit of large pressure, which will also generalize to hyperbolic space. We will explain this in the next section.

\vspace{10pt}

\noindent {\bf Validity of the flat-space approximation for hyperbolic space}

\noindent If the size of the self-avoiding loop is small compared to $\ell$, then we can approximate hyperbolic space by flat space, and the RGJ formula will give the answer for ordinary JT gravity. Let's estimate when this approximation is valid. The typical size of an unpressurized self-avoiding loop in flat space is of order $\upbeta^{3/4}$, so if the pressure is sufficiently small, the condition for applying the RGJ formula is $\upbeta^{3/4} \ll \ell$. On the other hand, if the pressure is large enough, the linear size of the loop is of order $(p\upbeta^3)^{1/2}$, as we explain in (\ref{rhosize}) below. A uniform condition for validity is therefore
\be
\max\{\upbeta^{3/4},(p\upbeta^3)^{1/2}\}\ll \ell.
\ee
This condition is enough to give qualitative accuracy of the RGJ formula. But in the large-pressure region, the free energy is very large, so even small changes can give a large multiplicative correction to $Z(\upbeta)$. As we will see below, for the multiplicative error to be small, we need a stronger condition
\be\label{flatspaceapprox}
\max\{\upbeta^{3/4},p^{3/2}\upbeta^3\}\ll \ell.
\ee
Note that for any value of $p$, we can find a sufficiently small $\upbeta$ so that this condition holds. This means that the behavior of $Z$ in the high-temperature limit is dominated by the flat space physics and the RGJ formula.\footnote{Actually this statement continues to hold for dilaton gravities with more general potential.}

\section{The intermediate regime}\label{sec:largePressure}
\subsection{Classical computations}
\begin{figure}[t]
\begin{center}
\includegraphics[width = .9\textwidth]{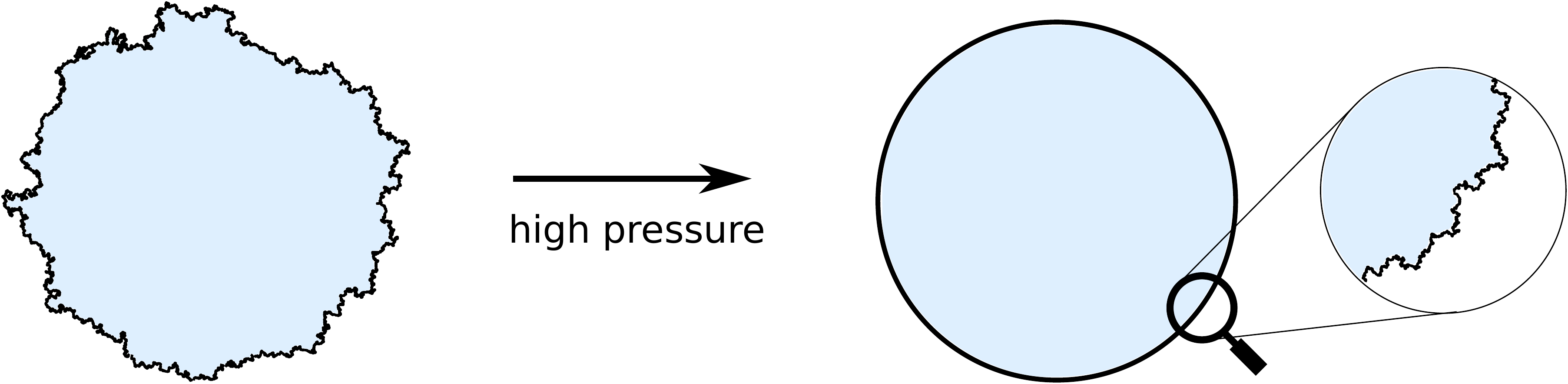}
\caption{{\small In the large pressure limit, the boundary will be macroscopically a circle at some radius $\rho$, but microscopically it will look like a typical self-avoiding walk.}}\label{fig:largePressure}
\end{center}
\end{figure}

For self-avoiding loops in either flat or curved space, one can use a simple effective theory to compute the partition function when the pressure $p$ is large. In this limit, on macroscopic scales a typical self-avoiding loop will resemble a smooth circle. On microscopic scales, however, it will be highly erratic, like a standard self-avoiding walk, see figure \ref{fig:largePressure}. The idea of the effective theory is to replace the microscopic details of the SAW by an entropic force that wants to pull the circle smaller. At some optimal size of the circle, this force will be balanced by the pressure.

The description will be accurate in the regime where the self-avoiding walk is stretched quite tight, so that there is a large separation between the scale where the boundary becomes wiggly and the scale of the circle.

\begin{figure}
\begin{center}
\includegraphics[width = .7\textwidth]{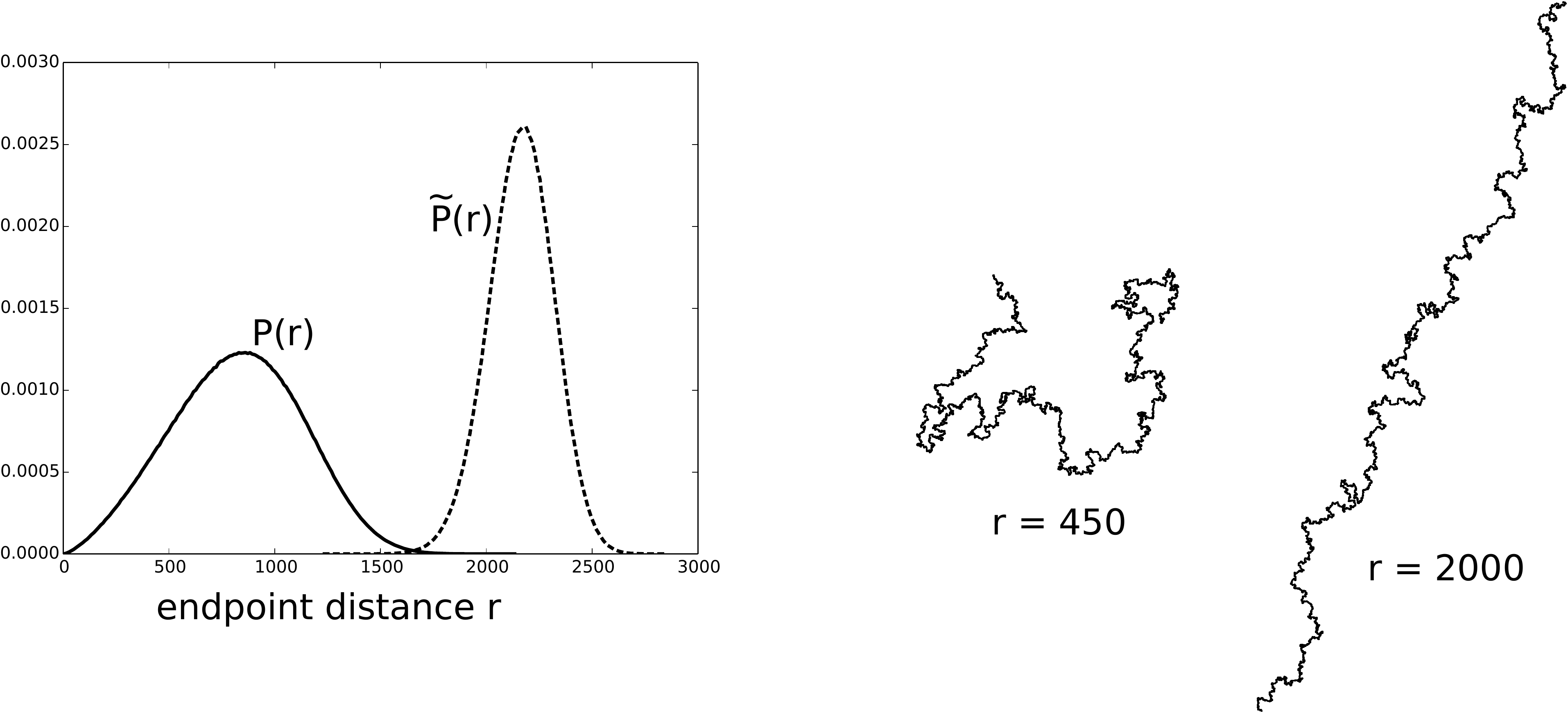}
\caption{{\small The probability distribution for the distance between the endpoints of a self-avoiding walk of $10^4$ steps of unit length is plotted (solid) together with the the reweighted probability distribution $\tilde{P}(r) \propto e^{Fr}P(r)$ when we include a force pulling the endpoints apart (dotted). Two walks representative of different parts of the probability distribution are shown. The distributions and walks were generated using Monte Carlo using \cite{kennedy2002faster}.}}\label{fig:endpointDistances}
\end{center}
\end{figure}

The main input to this effective theory is the resistance to stretching of a self-avoiding walk. Ideally, what we would like to know is the probability distribution for the distance between the endpoints. This can be evaluated using Monte Carlo (see figure \ref{fig:endpointDistances}), but no exact formula is known. However, since our goal is to work in the region of large pressure, it will be good enough to know the behavior of the tail of the probability distribution, when the walk is stretched out quite straight. This is conjectured to have the form \cite{domb1965shape,fisher1966shape,mckenzie1971shape}
\be\label{prob}
P(r) \propto r^{1 + \frac{5}{8}} \exp\left(-c_2\frac{r^4}{N^3 a^4}\right),
\ee
where $c_2$ is a non-universal constant that depends on the lattice. 

In the region where the SAW is stretched quite tight, the power-law prefactor will be ignorable relative to the exponential. In this approximation, the distribution has an important property of subdivision invariance. Suppose that we take a SAW of $N$ steps, and divide it into $k$ walks of $N/k$ steps. If each walk has length $l$, and they are oriented in the same direction, then the full walk will have length $L = kl$. Since
\be
\frac{L^4}{N^3} = k \frac{l^4}{(N/k)^3},
\ee
we get the same leading answer for the probability for the total length by applying (\ref{prob}) to the full walk or by applying it $k$ times to the subwalks.

Let's now apply this to the circle. We can think of the circle as built out of $k$ approximately straight segments. Applying the subdivision invariance argument, we find that for a given circle, the statistical factor is proportional to 
\be
\exp\left(-c_2\frac{L^4}{N^3a^4}\right).
\ee
where $L$ is the circumference of the circle. Substituting in (\ref{defupbeta}) and including the weighting from the area term, we find the leading approximation to the partition function
\begin{align}
\log\,Z(\upbeta) &\sim \max_{\text{circle size}}\left[p\cdot(A+2\pi \ell^2) - \frac{c_2}{c_1^3} \frac{L^4}{\upbeta^3}\right].\label{circ}
\end{align}
In order to compute the entropy $S(E)$, assuming $E <0$, we would like to evaluate
\begin{align}\label{tototo}
S(E) &= \text{ext}_{\upbeta,\text{ circle size}} \left[p\cdot(A+2\pi \ell^2) - \frac{c_2}{c_1^3} \frac{L^4}{\upbeta^3} + \upbeta E\right].
\end{align}

In this expression for a supposedly universal quantity, we find a ratio of non-universal coefficients $c_2/c_1^3$. For consistency, this ratio should be universal, and in fact we will find agreement with the RGJ formula only if it has the specific value
\be\label{c1111}
\frac{c_2}{c_1^3} = \frac{3}{(4\pi)^2}.
\ee
In appendix \ref{app:MC}, we describe a Monte Carlo estimate of $c_2$ on the square lattice. The result is consistent with existing results for $c_1$ \cite{jensen2000size} and this relationship, up to the three digits of precision that we were able to compute $c_2$. In what follows, we will assume that it is correct. 

Now, let's evaluate (\ref{tototo}). In hyperbolic space, the area of a circle at radius $\rho$ is $2\pi\ell^2[\cosh(\rho)-1]$, and its circumference is $2\pi\ell\sinh(\rho)$, where $\ell$ is the curvature length of the hyperbolic space. Extremizing over $\upbeta$ and $\rho$, and plugging in (\ref{c1111}), we find
\begin{align}\label{ent2}
S(E)=2\pi p\ell^2 \sqrt{1 - \frac{4}{3\pi p^2\ell^2}|E|^{3/2}}.
\end{align}
As $E$ approaches zero from below, the size of the optimal circle becomes small, and the hyperbolic answer reduces to $2\pi p \ell^2$ plus the answer we got from the RGJ formula (\ref{ent1}).

\subsection{One-loop computations}
To get a more accurate formula for $Z(\upbeta)$ at large pressure, we will now do the integral over small fluctuations around the classical configuration discussed in the previous section. The fluctuations we will integrate over are not the UV fluctuations of the microscopic self-avoiding walk; they are the fluctuations in the shape and parametrization of the smoothed self-avoiding walk, viewed on sufficiently long-distance scales that it is still pulled quite straight.

We will continue to use the following coordinates for hyperbolic space:
\be\label{hypmetric}
\mathrm{d}s^2 = \ell^2\big(\mathrm{d}\rho^2 + \sinh^2(\rho)\mathrm{d}\theta^2\big).
\ee
The flat-space answer can be obtained as a limit. The smoothed self-avoiding walk will be described as a parametrized curve $\{\rho(u),\theta(u)\}$, where $0\le u < 2\pi$. The parametrization is important, because the self-avoiding walk comes with a natural parametrization, proportional to the number of steps. We will expand around a circular configuration
\be
\rho(u) = \rho_0 + \delta\rho(u), \hspace{20pt}\theta(u) = u + \delta\theta(u).
\ee
Our first goal is to write an effective action to quadratic order for the fluctuations $\delta\rho$ and $\delta\theta$. The action has two pieces,
\be
I = I_{\text{area}} + I_{\text{entropic}}.
\ee
The first piece is, up to second order in the fluctuations,
\begin{align}
I_{\text{area}} &= -p(A +2\pi\ell^2)\\
 &= -\ell^2p\int_0^{2\pi}\mathrm{d}u \cosh(\rho(u))\theta'(u) \\
&= -\ell^2p\left\{2\pi \cosh(\rho_0) + \text{linear} + \int_0^{2\pi} \mathrm{d}u\left[\sinh(\rho_0)\delta\rho(u)\delta\theta'(u) + \frac{1}{2}\cosh(\rho_0)\delta\rho(u)^2\right]\right\}.\notag
\end{align}
We didn't write the piece linear in the fluctuations, because it will cancel when we impose that we are expanding about the ``optimal'' circle that minimizes the action.

The second piece in the action arises from integrating out the fluctuations in the SAW on microscopic scales. To derive this piece, we imagine dividing the path into $k$ segments, each made up of $N/k$ microscopic steps. We will choose $k$ so that the segments are small compared to the size of the circle and also small compared to the AdS scale $\ell$, but large enough that the SAW is approximately pulled straight on the scale of each segment. (Such a choice is possible in the large pressure limit, as we will discuss below.) Using the large $r$ form of the distribution (\ref{prob}), the entropic action is then
\be\label{entropic}
I_{\text{entropic}} = c_2\sum_{i = 1}^k \frac{(\Delta s_i)^2}{(N/k)^3a^4}.
\ee
where $\Delta s_i$ is the proper length of the $i$-th segment. We can rewrite $N/k$ in terms of the interval of our parameter $u$ using
\be
\Delta u = \frac{2\pi}{\upbeta}\frac{a^{4/3}N}{c_1k }.
\ee
Substituting this into (\ref{entropic}), using (\ref{c1111}) to evaluate $c_2/c_1^3$, and then approximating the sum over $i$ as an integral over $u$, we have
\begin{align}\label{entropic2}
I_{\text{entropic}} = \frac{3}{(4\pi)^2}\left(\frac{2\pi}{\upbeta}\right)^3\sum_{i = 1}^k\frac{(\Delta s_i)^4}{(\Delta u)^3}\approx \frac{3\pi}{2\upbeta^3}\int_0^{2\pi} \mathrm{d}u\left(\frac{\mathrm{d}s}{\mathrm{d}u}\right)^4.
\end{align}
Note that this expression depends on more than just the total length of the curve. It depends on the specific way that it is parametrized as a function of $u$. The geometrical quantity $\mathrm{d}s/\mathrm{d}u$ can be worked out using the metric (\ref{hypmetric}). We have
\begin{align}
\frac{1}{\ell}\frac{\mathrm{d}s}{\mathrm{d}u} &=\sqrt{ \sinh^2(\rho)(1+\delta\theta')^2  +(\delta\rho')^2}\\
&\approx \sinh(\rho)(1+\delta\theta') + \frac{1}{2}\frac{(\delta\rho')^2}{\sinh(\rho_0)},
\end{align}
where the expression on the second line is correct to quadratic order in the fluctuations. The integrand of (\ref{entropic2}) is then
\begin{align}
\frac{1}{\ell^4}\left(\frac{\mathrm{d}s}{\mathrm{d} u}\right)^4= \sinh^4(\rho_0)\left\{1 + \text{linear} + \left[2 + \frac{6}{\tanh^2(\rho_0)}\right](\delta\rho)^2 + 6(\delta\theta')^2 + \frac{16\,\delta\theta'\delta\rho}{\tanh(\rho_0)} + \frac{2(\delta\rho')^2}{\sinh^2(\rho_0)}\right\}.\notag
\end{align}

We have now computed all of the terms that will appear in the effective action. They can be simplified using the saddle point equation that relates the optimal value of $\rho_0$ to $\upbeta$. To work this out, we first write the full action at zeroth order in the fluctuations, by adding together the zeroth order terms in $I_{\text{area}}$ and $I_{\text{entropic}}$. This gives a quantity that we will refer to as $I_{0}$:
\begin{align}\label{classical1}
I_{0} &= -2\pi\ell^2p\cosh(\rho_0) + \frac{3\pi^2\ell^4}{\upbeta^3}\sinh^4(\rho_0).
\end{align}
The equation of motion that follows from extremizing this over $\rho_0$ is
\be\label{eom}
p\upbeta^3 = 6\pi \ell^2\sinh^2(\rho_0)\cosh(\rho_0).
\ee
Normally, one would solve this for $\rho_0$ and then substitute in to find the on-shell action as a function of the control parameter $\upbeta$. However, the formulas are much simpler if we write everything in terms of $\rho_0$, which we view implicitly as a function of $\upbeta$ by (\ref{eom}). In this way of doing things, we actually use (\ref{eom}) to eliminate $\upbeta$ from (\ref{classical1}). Then the on-shell action is
\begin{align}
I_{0}=-\frac{\pi}{2}\ell^2 p \left[3\cosh(\rho_0) + \cosh^{-1}(\rho_0)\right].
\end{align}
The saddle point condition (\ref{eom}) can be used to simplify the rest of the action too. The linear terms in the action will vanish, since we are expanding about a stationary point. And using (\ref{eom}), the quadratic terms from $I_{\text{area}}$ and $I_{\text{entropic}}$ can be combined to give a simple expression for the full action up to quadratic order:
\begin{align}
I = I_{0} + \frac{p\ell^2}{2}\int\mathrm{d}u\left\{\frac{(\delta\rho')^2 - (\delta\rho)^2}{\cosh(\rho_0)} + 3\frac{\sinh^2(\rho_0)}{\cosh(\rho_0)}\big[\coth(\rho_0)\delta\rho + \delta\theta'\big]^2\right\}.\label{action2}
\end{align}

To get the one-loop expression for $Z(\upbeta)$, we now just need to do the functional integral over $\delta\rho$ and $\delta\theta$. To do so, it is convenient to decompose $\delta\rho$ and $\delta\theta$ into Fourier modes:
\be
\delta\rho(u) = \delta\rho_0 + \sum_{n\neq 0} e^{-\mathrm{i}nu}\left[ \rho^R_n + \mathrm{i}\rho^I_n\right], \hspace{20pt} \delta\theta(u) = \theta_0 + \sum_{n\neq 0} e^{-\mathrm{i}nu}\left[ \theta^R_n + \mathrm{i}\theta^I_n\right].
\ee
Inserting these expansions, the quadratic action for the $n\neq 0$ modes in (\ref{action2}) becomes
\begin{align}
\frac{2\pi p\ell^2}{\cosh(\rho_0)} \sum_{n = 1}^\infty \left\{(n^2{-}1)\big[(\rho^R_n)^2{+}(\rho_n^I)^2\big] + 3\sinh^2(\rho_0)\left[\big(\coth(\rho_0)\rho^R_n {+} n \theta^I_n\big)^2+\big(\coth(\rho_0)\rho^I_n {-} n \theta^R_n\big)^2\right]\right\},\notag
\end{align}
and for the $\delta\rho_0$ mode it becomes:
\be
\pi p\ell^2(3\cosh\rho_0-\cosh\rho_0^{-1})\delta\rho_0^2.
\ee
An important feature of the action is that it has three zero modes, corresponding to the parameters $\epsilon_0,\epsilon_1,\epsilon_2$ in
\be
\delta\rho = \epsilon_1 \cos(u) + \epsilon_2\sin(u), \hspace{20pt} \delta\theta = \epsilon_0 - \epsilon_1 \coth(\rho_0)\sin(u) + \epsilon_2\coth(\rho_0)\cos(u).
\ee
These three correspond to rotations anad translations of the JT gravity region within the parent hyperbolic space. Since these transformations do not change any intrinsic feature of the JT gravity region, they must be regarded as pure gauge. To fix these modes, we would like to insert
\be
\frac{\lambda^2}{\ell^2}\delta(\epsilon_1)\delta(\epsilon_2)\delta(\epsilon_0)
\ee
inside the path integral. Here $\lambda$ is an arbitrary length scale that arises when we gauge-fix the spatial translations. In principle, there is another arbitrary factor in the gauge-fixing for rotations, but we will use the convention where the volume of the group of rotations is $2\pi$. 

To implement this gauge-fixing, we can include in the path integral the insertion
\be
\frac{\lambda^2}{\ell^2}\times\delta(\rho_1^R) \frac{\partial \rho_1^R}{\partial\epsilon_1}\times\delta(\rho_1^I)\frac{\partial \rho_1^I}{\partial\epsilon_2}\times \delta(\theta_0)\frac{\partial\theta_0}{\partial\epsilon_0} = \frac{\lambda^2}{4\ell^2}\delta(\rho_1^R)\delta(\rho_1^I)\delta(\theta_0).
\ee
Finally, we can now do the Gaussian integrals, with the ultralocal measure\footnote{It seems like we have neglected many multiplicative factors in our derivation. For example, in writing the entropic part of the action, we included the exponential term in (\ref{prob}), but not the prefactor. Similarly, in writing the measure, we neglected the fact that the correct discretized measure will have a volume element factor $\sinh(\rho_i)\mathrm{d}\theta_i\mathrm{d}\rho_i$ at each point. In the large pressure approximation, such factors will be constant, because fluctuations in $\rho$ and $\theta$ are small. But there will still be an unknown multiplicative constant multiplying the path integral at each of the $k$ points in the discretization. Fortunately, a product of a constant factor at each point can be entirely absorbed into a length counterterm, so such factors actually do not affect the normalization of the final answer.}
\be
\left(\mathrm{d}\delta\rho_0 \prod_{n = 1}^\infty 2\mathrm{d}\rho^R_n\mathrm{d}\rho^I_n\right)\left(\mathrm{d}\theta_0\prod_{n = 1}^\infty 2\mathrm{d}\theta^R_n\mathrm{d}\theta^I_n\right).
\ee
This leads to the un-simplified formula
\begin{align}
Z(\upbeta) &= e^{-I_{0}}\frac{2\lambda^2}{4\ell^2}\left(\frac{\pi}{\pi p \ell^2(3\cosh(\rho_0) - \cosh^{-1}(\rho_0))}\right)^{1/2}\prod_{n = 2}^\infty 2\frac{\pi\cosh(\rho_0)}{2\pi p\ell^2(n^2-1)}\prod_{n = 1}^\infty 2\frac{\pi\cosh(\rho_0)}{6\pi p \ell^2\sinh^2(\rho_0)n^2}.\notag
\end{align}
The products can be simplified and then evaluated using
\be
\prod_{n = 1}^\infty c \rightarrow c^{-1/2}, \hspace{20pt} \prod_{n = 1}^\infty n^2 \rightarrow 2\pi, \hspace{20pt} \prod_{n = 2}^\infty (1-n^{-2}) = \frac{1}{2}.
\ee
The arrows indicate the answer after regularization and renormalization, or after using e.g.~zeta function regularization. Using these formulas, the answer for $Z(\upbeta)$ reduces to
\begin{align}\label{oneloop}
Z(\upbeta) = \frac{\lambda^2}{4\pi^2\ell^2}\left(\frac{ \sinh^2(\rho_0)}{\cosh^2(\rho_0) - \frac{1}{3}}\right)^{1/2}\hspace{-5pt}\left(\frac{p\ell^2}{\cosh(\rho_0)}\right)^{3/2}\exp\left\{\frac{\pi}{2}\ell^2 p \left[3\cosh(\rho_0) + \cosh^{-1}(\rho_0)\right]\right\}.
\end{align}

In this expression, $\rho_0$ should be understood as a function of $\upbeta$ using (\ref{eom}). For general values of $\upbeta$, it is not easy to solve for $\rho_0$ explicitly. But in two extreme limits, it can be solved simply:
\be
\rho_0 = \begin{cases}\left(\frac{p\beta^3}{6\pi \ell^2}\right)^{1/2} & p\upbeta^3 \ll \ell^2 \\\log\left[\left(\frac{4p\beta^3}{3\pi \ell^2}\right)^{1/3} + \frac{1}{3}\left(\frac{4p\beta^3}{3\pi \ell^2}\right)^{-1/3}\right] & p\upbeta^3 \gg \ell^2. \end{cases}\label{rhosize}
\ee
Plugging these expressions into (\ref{oneloop}) and taking the appropriate limits of the trigonometric functions, we find
\be
Z(\upbeta) \approx \begin{cases}\frac{\lambda^2}{8\pi^{5/2}}p^2\upbeta^{3/2}\exp\left[p\ell^2\left(2\pi + \frac{p\upbeta^3}{12\ell^2} + O(\frac{p^2\upbeta^6}{\ell^4})\right)\right] & p\upbeta^3 \ll \ell^2 \\ 
\frac{3^{1/2}\lambda^2}{(2\pi)^{3/2}}\frac{p\ell^2}{\upbeta^{3/2}}\exp\left[p\ell^2\left(\#\upbeta + \left(\frac{6\pi^4\ell^2}{p\upbeta^3}\right)^{1/3} + O(\frac{\ell^2}{p\upbeta^3})\right)\right] & p\upbeta^3 \gg \ell^2.\label{limitsOfAnswer}
\end{cases}
\ee
Note that for the corrections to be small in absolute terms (as opposed to just small compared to the terms shown) we need the stronger conditions $p \ll \ell^{2/3}/\upbeta^2$ for the first line to be accurate, and $\upbeta \gg \ell^{4/3}$ for the second line to be accurate.

A pleasant surprise is that the one-loop density of states $\density(E)$ corresponding to (\ref{oneloop}) is very simple. To compute it, one can do the inverse Laplace transform
\be\label{ilt}
\density(E) = \int_{\text{const} + \mathrm{i}\mathbb{R}}\frac{\mathrm{d}\upbeta}{2\pi \mathrm{i}}e^{\upbeta E}Z(\upbeta)
\ee
of (\ref{oneloop}) in the saddle-point approximation. One finds
\be
\density(E) = \frac{\lambda^2}{4\pi^3}\sqrt{|E|}\exp\left[2\pi p \ell^2\sqrt{1 - \Big(\frac{E}{E_0}\Big)^{3/2}}\right], \hspace{20pt} E_0 = -\left(\frac{3\pi\ell^2p^2}{4}\right)^{2/3}.
\ee
This is the formula that was reported in the Introduction in (\ref{densInt}).

\vspace{10pt} 

\noindent{\bf Validity of the large-pressure approximation}

\noindent There are two conditions for the validity of the large-pressure approximation that we used in this section. First, to have a local effective theory, we have to be able to ignore the nonlocal self-avoiding constraint at the distance scales where the theory applies. This will be allowable if the self-avoiding walk is pulled approximately straight down to a scale that is much smaller than both the size of the circle and $\ell$. 

The length scale where UV wiggles appear in the self-avoiding walk, $L_{\text{wiggles}}$, can be estimated as follows. If we zoom in on a small piece of the pressurized self-avoiding loop, it will locally be described as a self-avoiding walk with some tension pulling the endpoints apart. This tension is the only parameter that is important for the local properties of the loop, and dimensional analysis implies that the lengthscale $L_{\text{wiggles}}$ should be the inverse of this tension. We can work out the tension by taking the derivative of the pressure term with respect to the total length, finding
\be
\frac{1}{L_{\text{wiggles}}} \sim \frac{\partial}{\partial L}p A \sim p\tanh(\rho_0)\ell.
\ee
For the effective description to be possible, we need $L_{\text{wiggles}}\ll \ell$ and also $L_{\text{wiggles}}\ll \sinh(\rho_0)\ell$. Both conditions will be satisfied if $
p\ell^2 \gg 1$ and $p\ell^2\rho_0^2\gg 1$. Using (\ref{rhosize}), one can rewrite these conditions in terms of $\upbeta$ as
\be\label{effective}
p \ell^2\gg 1, \hspace{20pt} p\upbeta^{3/2}\gg 1. 
\ee

A second constraint for the validity of our approximations is that the effective theory should be weakly coupled. For this, we need the fluctuations $\delta\rho$ and $\delta\theta$ to be small compared to one. This amounts to checking that the coefficients of both terms in the action (\ref{action2}) are large. Using (\ref{rhosize}), this requirement reduces to
\be\label{effective2}
p\ell^{4/3}\gg \upbeta^{3/2},\hspace{20pt} p\upbeta^{3/2}\gg 1.
\ee
As it turns out, these conditions are stronger than (\ref{effective}).

\vspace{10pt}

\noindent {\bf Agreement with the flat space answer in region of overlap}

\noindent The conditions (\ref{effective2}) overlap with the conditions for the validity of the flat space approximation (\ref{flatspaceapprox}), and we can check for agreement in the region of overlap. Using the large argument behavior of the RGJ function $f(y)$ in (\ref{flargey}), and plugging it into (\ref{rhrrjrhrj}), we find the large-pressure limit of the flat space answer
\be
Z(\upbeta) \approx \frac{\lambda^2}{8\pi^{5/2}}p^2\upbeta^{3/2}\exp\left(\frac{p^2\upbeta^3}{12}\right).
\ee
This agrees with the flat space limit of the large pressure answer (\ref{limitsOfAnswer}) in the $p\upbeta ^3\ll \ell^2$ regime, after multiplying by $e^{2\pi p\ell^2}$ to account for the extra factor in (\ref{proposal2}) compared to (\ref{proposal}). In order for the ratio of the two answers to be close to one, we need the correction term in (\ref{limitsOfAnswer}) to be small, which requires $p\ll \ell^{2/3}/\upbeta^2$.

\section{The Schwarzian regime}
In the standard treatment of JT gravity with asymptotic boundary conditions, the statistical mechanics of the shape of the boundary is described by the Schwarzian theory. This theory captures smooth long-distance fluctuations on scales larger than $\ell$. We can ask how to recover this description starting from the self-avoiding walk at finite size and pressure.

One necessary condition is that the pressure should be large compared to $\ell^{-2}$,
\be
p\ell^2 \gg 1.
\ee
In this situation, the self-avoiding walk will be pulled approximately straight down to scales of order $1/(p\ell) \ll \ell$. The microscopic wiggles can be integrated out on scales smaller than this, leading to an effective theory describing smooth long-distance fluctuations on scales of order $1/(p\ell)$ and larger. 

The other necessary condition is that the total size of the loop should be much larger than $\ell$. From (\ref{rhosize}), we see that this will be true if
\be\label{cond2Sch}
p \upbeta^3 \gg \ell^2.
\ee
This is enough for qualitative accuracy of the Schwarzian theory, but for $Z(\upbeta)$ to be correct up to small multiplicative errors, we will see that this last condition needs to be replaced by the stronger condition
\be
\upbeta \gg \ell^{4/3}.
\ee

Now, to understand how one can recover the Schwarzian theory in this limit, it is helpful to look at the quadratic action for small fluctuations in the effective theory (\ref{action2}). When (\ref{cond2Sch}) is satisfied, we will have $\rho_0 \gg 1$, and then the action becomes, schematically
\be\label{kwjwkw}
I_{\text{quad}} \sim p\ell^2\int_0^{2\pi} \mathrm{d}u\left\{\frac{(\delta \rho')^2 - (\delta \rho)^2}{e^{\rho_0}} + e^{\rho_0}\big[\delta\rho + \delta\theta'\big]^2\right\}.
\ee
Since $\rho_0$ is large, the coefficient of the second term is large, freezing $\delta\rho + \delta \theta' = 0$. This means that the fluctuations in the parametrization and length of the cuve disappear, and the only remaining fluctuations are in its shape. Substituting $\delta\rho = -\delta \theta'$ into the first term in (\ref{kwjwkw}), we get the quadratic approximation to the Schwarzian theory. However, because the coefficient of the first term can be small (depending on the ratio of $p$ and $e^{\rho_0}$), these shape fluctuations can be large, and the quadratic approximation we made may not be valid. Instead, the fluctuations are expected to be described by the nonlinear Schwarzian theory.

Note that in this regime, the microscopic details of the walk (i.e.~the self-avoiding constraint) should not matter, since the important fluctuations are on large scales. And in the case where the walk is an ordinary random walk, \cite{Kitaev:2018wpr,Yang:2018gdb} showed by explicit calculation that one does recover the results of the nonlinear Schwarzian theory in the analogous limit.

\vspace{10pt}

\noindent {\bf Agreement with the large pressure theory in region of overlap}

\noindent The partition function in the Schwarzian theory with the convention that $\lambda = \ell$ is\footnote{The partition function of the Schwarzian theory has been worked out in \cite{Cotler:2016fpe,Stanford:2017thb,Bagrets:2017pwq,Mertens:2017mtv,Belokurov:2017eit}. The numerical prefactor with the convention $\lambda = \ell$ was computed in \cite{Kitaev:2018wpr,Yang:2018gdb,Saad:2019lba}.}
\be\label{schanswer}
Z(L) = \frac{1}{(2\pi)^{1/2}}\left(\frac{p\ell^3}{L}\right)^{3/2} e^{2\pi^2 \frac{p\ell^3}{L}},
\ee
where $L$ is the length of the smooth boundary curve. To relate this to our large-pressure theory, we need to related $\upbeta$ and $L$. In the Schwarzian theory, $L$ is the fixed length of the boundary curve. We should compare this to the length of the smoothed curve that appears in the effective theory. The saddle point relation (\ref{eom}) determines this length in terms of $\upbeta$, which implies
\be
L \rightarrow 2\pi\ell\sinh(\rho_0) \approx \left(\frac{(2\pi)^2p\ell}{3}\right)^{1/3}\upbeta.
\ee
Plugging this into (\ref{schanswer}), and using the value $\lambda = \ell$, we find agreement with (\ref{limitsOfAnswer}) in the $p\upbeta^3 \gg \ell^2$ regime, after adjusting a counterterm to remove the $\# \upbeta$ term in (\ref{limitsOfAnswer}). For the ratio of the two answers to be close to one, we need the first subleading correction in (\ref{limitsOfAnswer}) to be small, which requires the stronger condition $\upbeta \gg \ell^{4/3}$.

\section{Discussion}
Our motivation for studying JT gravity at finite cutoff was an interesting discrepancy between the random-walk results of \cite{Kitaev:2018wpr,Yang:2018gdb} and classical computations in JT gravity. In this Discussion, we will make some comments on the discrepancy.

First, let's review the classical approximation for the partition function of JT gravity. This is a saddle-point treatment of the path integral, where we replace the integral $\mathcal{D}\gamma$ by the single loop $\gamma$ that maximizes the enclosed area for a given length $L$. This will be a circle, and we can take it to be the constant $\rho$ locus, where $L=2\pi \ell\sinh\rho$. The area enclosed is $2\pi\ell^2(\cosh(\rho) - 1)$, and the leading approximation to the partition function is
\be
\log\,Z(L) \approx -I_{JT} =  2\pi p\ell^2\cosh \rho = 2\pi p\ell^2\sqrt{1+\Big({L\over 2\pi\ell}\Big)^2}.
\ee
If we interpret this as a thermal partition function with inverse temperature $L$, then the corresponding entropy as a function of the conjugate energy $E_L$ is
\be\label{entropyyyyy}
S(E_L) = \text{ext}_L \Big[L E_L + \log\,Z(L)\Big] = 2\pi p\ell^2\sqrt{1 - \left(\frac{E_L}{p\ell}\right)^2}.
\ee

Interestingly, this result for the entropy is consistent with the conjecture that a 1d version of the $T\bar{T}$ deformation describes a finite cutoff in JT gravity \cite{Gross:2019ach,Gross:2019uxi,TTBAR}. However, it differs from the large-pressure limit of the ordinary random walk results of \cite{Kitaev:2018wpr,Yang:2018gdb} and from the self-avoiding walk results in this paper (\ref{ent2}). After choosing the ground state $E_0$ appropriately, all three answers can be written in the form
\be
S(E) = 2\pi p \ell^2\sqrt{1 - \left(\frac{E}{E_0}\right)^{2\nu}}
\ee
where $\nu = 1$ for classical JT gravity, $\nu = 1/2$ for the ordinary random walk, and $\nu = 3/4$ for the self-avoiding case. In the standard asymptotic limit of JT gravity, the energy is scaled near the ground state, so that we have $S\propto \sqrt{E-E_0}$ in all three cases. But at finite cutoff, the answers are meaningfully different.

\subsection{A free particle analogy}
To understand this difference, we found the following analogy helpful. Consider the relativistic propagator of a massive particle in two dimensions,
\be\label{toMatch}
G(x) = \int \frac{\mathrm{d}^2p}{(2\pi)^2} \frac{e^{\mathrm{i}p\cdot x}}{p^2+m^2} = \frac{1}{2\pi}K_0(m|x|).
\ee
It is sometimes useful to think about this propagator as being a sum over all worldline paths, weighted by an action given by the length of the path:
\be\label{intrep}
G(x) \sim \int_{\gamma(0) = 0}^{\gamma(1) = x} \mathcal{D}\gamma \, e^{-\mu\, L(\gamma)}.
\ee
There are actually two rather different senses in which this is true, depending on whether we view this integral as an effective description (IR theory) or as an exact one (UV theory).

First, we can take this integral seriously as an exact description \cite{Cohen:1985sm,polyakov1987gauge,Kitaev:2018wpr}. Typical paths will be microscopically fractal, but one can regularize them on e.g.~a square lattice of size $a$, and define a renormalized version of the length $L\rightarrow \upbeta = La/2$. The propagator for a fixed value of this renormalized length is the familiar propagator of an ordinary random walk, proportional to $e^{-x^2/(2\upbeta)} / \upbeta$. Inserting this answer and doing the integral over the length at the end, we get
\be\label{quantum}
G(x) \propto \int_0^\infty \frac{\mathrm{d}\upbeta}{\upbeta} \ \exp\left(-\mu \upbeta - \frac{x^2}{2\upbeta}\right) \propto K_0(\sqrt{2\mu}|x|).
\ee
This reproduces the correct answer (\ref{toMatch}) for an appropriate choice of $\mu$.

Second, we can view (\ref{intrep}) as an effective theory, appropriate for large mass and/or large $x$. At leading order, we simply evaluate the action on the classical configuration, which is a straight path from the origin to $x$. This leads to
\be
G(x) \sim e^{-\mu |x|}.
\ee
Here we find an apparent contradiction between the two approaches. For large $|x|$, the classical answer decays as $e^{-\mu|x|}$ and the quantum answer (\ref{quantum}) decays as $e^{-\sqrt{2\mu}|x|}$. These answers are analogous to the different formulas for the entropy that we found above. However, for the particle we can reconcile the answers by defining
\be\label{murelRW}
\mu_L = \sqrt{2\mu_\upbeta} = m.
\ee
Here $\mu_\upbeta$ is conjugate to the renormalized microscopic length $\upbeta$ in the exact theory, and $\mu_L$ is conjugate to the smooth length $L$ in the effective description. With this understanding, the theories are consistent at large $|x|$. One can try to improve the accuracy of the effective description, by integrating over small fluctuations around the straight classical path. The one loop computation makes sense and gives the correct coefficient, but higher orders in the perturbation theory give UV divergences, and one has to resort back to the exact description.

One can derive the IR effective description from the UV description by writing the random walk propagator with fixed $\upbeta$ in terms of a Gaussian functional integral, and then replacing the integral over $\upbeta$ by an integral over a one-dimensional metric $e$ \cite{Cohen:1985sm,polyakov1987gauge}. Finally, by doing a saddle-point approximation for the integral over $e$, we get the effective theory with the relationship (\ref{murelRW}):
\be\label{manip2}
G \propto\int_0^{\infty}\mathrm{d}\upbeta \int\mathcal{D} x e^{-\int_0^{\upbeta} {1\over 2}\dot x^2+\mu_{\upbeta}}\propto\int {\mathcal{D}e \mathcal{D}x\over \text{Diff}}e^{-\int {1\over 2 e}\dot x^2+\mu_{\upbeta} e}\sim\int{\mathcal{D}x\over \text{Diff}} e^{-\sqrt{2\mu_{\beta}}\int |\dot x|}.
\ee

\subsection{JT gravity as an effective description of JT gravity}
In this paper, we have viewed JT gravity as an exact theory, and we have analyzed the path integral in the spirit of the first description of (\ref{intrep}) above. However, one can make contact with JT gravity viewed as an effective description by starting with the large-pressure effective theory of section \ref{sec:largePressure}. Integrating over $\upbeta$ to go to the microcanonical ensemble, we have
\begin{align}
\density(E_\upbeta) &\sim \int \mathrm{d}\upbeta\, e^{E_\upbeta \upbeta}\int \mathcal{D}x\, e^{-I_{\text{area}} - I_{\text{entropic}}} = \int \mathrm{d}\upbeta \int \mathcal{D}x\, e^{-I_{\text{area}} -\int_0^\upbeta (\frac{3}{(4\pi)^2}\dot{x}^4 -E_\upbeta)}.
\end{align}
In the second step, we used the formula for $I_{\text{entropic}}$ in (\ref{entropic2}). Following the steps in (\ref{manip2}), we now replace the integral over $\upbeta$ by an integral over a metric $e$ modulo reparameterizations:
\begin{align}
\density(E_\upbeta)\sim \int \frac{\mathcal{D}e\mathcal{D}x}{\text{Diff}} e^{-I_{\text{area}}-\int (\frac{3}{(4\pi)^2 e^3}\dot{x}^4 - E_\upbeta e)}\sim \int\frac{\mathcal{D}x}{\text{Diff}} e^{-I_{\text{area}}-\sqrt{\frac{4}{3\pi}}(-E_\upbeta)^{3/4}\int |\dot{x}|}.
\end{align}
To get the final expression, we integrated out the metric $e$ in a saddle point approximation. The result can be trivially rewritten as
\begin{align}
\density(E_\upbeta) \sim \int\frac{\mathcal{D}x}{\text{Diff}} e^{-I_{\text{area}}+E_L L}\label{jtjt}
\end{align}
where $L$ is the smoothed length of the curve in the large-pressure effective theory, and where the energy $E_L$ is defined by
\be\label{ELEB}
-E_L = \sqrt{\frac{4}{3\pi}}(-E_\upbeta)^{3/4}.
\ee

The point to notice is that (\ref{jtjt}) is the standard JT gravity formula for the partition function in the microcanonical ensemble, with energy $E_L$. So the effective description of JT gravity at large pressure is again JT gravity. However, the energies in the two descriptions are related nonlinearly, as with the $\mu$ parameters in the particle analogy. This is enough to explain the difference between the classical and quantum formulas for the entropy: substituting (\ref{ELEB}) into the classical formula for the entropy (\ref{entropyyyyy}) gives the answer (\ref{ent2}) from the large-pressure computation.

To summarize, the classical and quantum answers for the entropy disagree because they are defined as functions of energies that are conjugate to two different notions of length:
\begin{enumerate}
\item {\bf The renormalized UV length $\upbeta$:} In the quantum theory, typical trajectories of the boundary are fractal, but they can be given a renormalized notion of length, by regularizing the problem with some cutoff, and then multiplying the regularized length by a local renormalization factor. Here ``local'' means that the coefficient is allowed to depend on the microscopic details, but not on the total length or the pressure.

\item {\bf The IR length $L$:} In the large-pressure limit, there is another notion of length, which arises due to the fact that typical configurations become smooth on long distance scales, with a length that is well-defined as long as we don't zoom in too close. 
\end{enumerate}

\noindent Using (\ref{eom}), one finds that at large pressure, the two are related by
\be
\upbeta^3 = \frac{3}{4\pi^2p\ell}L^2\sqrt{L^2 + (2\pi\ell)^2}.
\ee
Normally, in JT gravity the length is taken to infinity, so that $L$ and $\upbeta$ are proportional and essentially equivalent. But for finite length they are nonlinearly related.

In this paper, we focused on $\upbeta$. The reason is that it isn't clear how to define $L$ at finite pressure. Relatedly, while the effective JT theory makes sense at the classical and one-loop level, at higher orders we expect UV divergences that require some input from a UV completion, such as the exact JT gravity theory defined in terms of $\upbeta$. 

One could try to define $E_L$ in general through the relationship (\ref{ELEB}) and then use the results of the UV theory. But because of the nonanalyticity in this relationship (\ref{ELEB}) at zero energy, the density of states $\density$ will be nonanalytic at $E_L = 0$, and unclear for $E_L>0$. There is a vaguely similar difficulty at high energy in \cite{TTBAR} and in the related $T\bar T$ discussion \cite{Gross:2019ach,Gross:2019uxi}. However, at least in the present context, it seems that the singularity at $E_L = 0$ is an artifact, since in terms of $E_\upbeta$ and (\ref{airyDensity}), nothing special happens at this point (see figure \ref{fig:airy}), and the spectrum continues smoothly up to infinite energy. This was also true of the random-walk analysis in \cite{Kitaev:2018wpr,Yang:2018gdb}.


\subsection*{Acknowledgments}
We are grateful to John Cardy, Luca Iliesiu, Jorrit Kruthoff, Eva Silverstein and Joaquin Turiaci for helpful discussions, and to the authors of \cite{TTBAR} for sharing a draft of their paper. ZY is supported in part by the Simons Foundation. This research was supported in part by the National Science Foundation under Grant No. PHY-1748958.

\appendix

\section{Deriving the density of states from the RGJ formula}\label{app:RGJ}
The purpose of this appendix is to derive the equation
\be\label{app:z}
\frac{1}{\beta^{3/2}}\sum_{n = 0}^\infty \frac{f_n}{\Gamma(\frac{3n-1}{2})}\beta^{3n/2} = \frac{1}{\pi^2}\int_{-\infty}^\infty \mathrm{d}E\, \frac{e^{-\beta x}}{\Ai^2(-E) + \Bi^2(-E)},
\ee
where the $f_n$ coefficients are given in (\ref{fn}). We start by defining $Z(\beta)$ as the LHS of (\ref{app:z}). Then we would like to write 
\be
Z(\beta) = \int_{-\infty}^\infty \mathrm{d}E\,\density(E) e^{-\beta E}\hspace{20pt} \iff \hspace{20pt} \density(E) = \int_{c + \mathrm{i}\mathbb{R}} \frac{\mathrm{d}\beta}{2\pi \mathrm{i}} e^{\beta E}Z(\beta)
\ee
and show that $\density(E)$ is the function appearing in the RHS of (\ref{app:z}). To carry this out, one can try to apply the inverse Laplace transform formula
\be
\int_{c + \mathrm{i}\mathbb{R}}\frac{\d \beta}{2\pi \i} e^{\beta E} \beta^a = \frac{1}{\Gamma(-a)E^{1+a}},
\ee
term-by-term in the sum on the LHS of (\ref{app:z}). This leads to a formal expression
\be\label{app:asymp}
\density(E) \sim \sum_{n = 0}^\infty \frac{f_n}{\Gamma(\frac{3n-1}{2})\Gamma(\frac{3(1-n)}{2})}\frac{\sqrt{E}}{E^{3n/2}} = -\sum_{n = 0}^\infty f_n\frac{\cos(\frac{3\pi n}{2})}{\pi}\frac{\sqrt{E}}{E^{3n/2}} = -\sum_{k = 0}^\infty f_{2k}\frac{(-1)^{k}}{\pi} \frac{\sqrt{E}}{E^{3k}}.
\ee
In going to the final expression, we noticed that the terms with odd $n$ drop out, and we relabeled $n \rightarrow 2k$. We refer to this as a formal expression because the factorial growth of $f_n$ for large $n$ \cite{kearney2009airy} implies that it is an asymptotic series.

We can make sense of the series by inserting the integral representation \cite{kearney2009airy}
\be
f_{2k} = -\frac{3}{4\pi^2}\int_\subset \mathrm{d}x \frac{x^{3k}}{\Ai^2(x) + \Bi^2(x)}\frac{1}{x^{3/2}}
\ee
where $x^{3/2}$ is defined to have a branch cut along the positive axis, and to have phase $-i$ along the negative real axis. The contour $\subset$ is defined to surround this cut in a clockwise direction. Inside the integral representation, the sum over $k$ is just a geometric series, and we find
\be\label{app:newint}
\density(E) = -\frac{3}{4\pi^3}\int_{\mathcal{C}} \mathrm{d}x \frac{1}{\Ai^2(x) + \Bi^2(x)} \frac{E^3}{E^3 + x^3}\frac{\sqrt{E}}{x^{3/2}}.
\ee
In this expression, we need to be careful with the choice of contour $\mathcal{C}$. The fact that the sum (\ref{app:asymp}) was asymptotic is reflected in the existence of poles in the integrand at the three roots of the equation $x^3 = E^3$, and different ways of routing the contour around these poles correspond to different completions of (\ref{app:asymp}). We will choose a completion such that the contour agrees with the original contour $\mathcal{C} = \subset$ in the case where $E$ is real and positive, so that the original series is alternating. We will then analytically continue the answer to all values of $E$.  (This is a standard practice in Borel resummation, but we are not trying to justify it rigorously because we know from numerical checks that it is correct.)

The contour $\subset$ can be deformed ``to the left'' and off to infinity, so that the integral becomes a sum of various residues of poles. First, we have poles where $\Ai^2(x) + \Bi^2(x)$ vanishes. These turn out to cancel. One can show this by using 
\be
\Ai(e^{\pm 2\pi \mathrm{i}/3}x ) = \frac{1}{2}e^{\pm \pi \mathrm{i}/3}\left(\Ai(x)\mp \mathrm{i}\Bi(x)\right), \hspace{20pt} \Ai(x)\Bi'(x) - \Ai'(x)\Bi(x) = \frac{1}{\pi}
\ee
to write a useful expression \cite{kearney2009airy}
\begin{align}
\frac{1}{\pi}\frac{1}{\Ai^2(x) + \Bi^2(x)}  &= \frac{\Ai(x)\Bi'(x) - \Ai'(x)\Bi(x)}{\Ai^2(x) + \Bi^2(x)} = \frac{1}{2\mathrm{i}}\left[\frac{\Ai'(x) + \mathrm{i}\Bi'(x)}{\Ai(x) + \mathrm{i}\Bi(x)} - \frac{\Ai'(x) - \mathrm{i}\Bi'(x)}{\Ai(x) - \mathrm{i}\Bi(x)}\right]\notag\\
&= \frac{1}{2\mathrm{i}}\left[e^{-2\pi \mathrm{i}/3}\frac{\Ai'(e^{- 2\pi \mathrm{i}/3}x)}{\Ai(e^{- 2\pi \mathrm{i}/3}x)}-e^{2\pi \mathrm{i}/3}\frac{\Ai'(e^{ 2\pi \mathrm{i}/3}x)}{\Ai(e^{ 2\pi \mathrm{i}/3}x)}\right].\label{app:id}
\end{align}
The Airy function has no poles, and it has zeros only along the negative real axis. So the poles in the two terms are at corresponding points along the rays $x \sim e^{\pm\pi \mathrm{i}/3}$. The rest of the integrand in (\ref{app:newint}) is equal along these two rays, so the residues of the poles are equal and opposite.

This cancellation implies that the full answer for the integral comes from the sum over the three remaining poles, coming from the three roots of the equation $x^3 + E^3 = 0$. The sum of these gives
\be
\density(E) = \frac{1}{\pi^2}\frac{1}{\Ai^2(-E) + \Bi^2(-E)}.
\ee
One half of this answer comes from the pole at $x = -E$. The other half comes from the sum of the two contributions at $x = e^{\pm \pi \mathrm{i}/3}E$, after using (\ref{app:id}) twice to simplify the sum.

\section{Large argument asymptotics of the RGJ formula}
Using the result of the previous section, we have that the $f(y)$ defined in (\ref{f}) has an integral representation
\be\label{app:Gairy}
f(y) = 2\sqrt{\pi}\frac{y}{\pi^2} \int_{-\infty}^\infty \mathrm{d}x \frac{e^{y^{2/3}x}}{\Ai^2(x) + \Bi^2(x)}.
\ee
We can use this to determine the large $y$ asymptotics. In the limit of large $y$, the exponential factor in the integrand favors large positive $x$, where we have the approximation
\be
\Ai^2(x) + \Bi^2(x) \approx \Bi^2(x) \approx \frac{1}{\pi x^{1/2}}e^{4x^{3/2}/3}.
\ee
Inserting this into (\ref{app:Gairy}), and approximating the integral using the saddle point approximation, we find
\be\label{flargey}
f(y) = e^{y^2/12}\left[y^2 + O(y^{-2})\right].
\ee

\section{Monte Carlo estimation of \texorpdfstring{$c_2$}{c2}}\label{app:MC}
In the main text of the paper, we found that agreement between the RGJ formula and the effective description requires a relationship
\be\label{c2c1app}
\frac{c_2}{c_1^3} = \frac{3}{16\pi^2}
\ee
between two non-universal coefficients $c_1$ and $c_2$. In this appendix, we will report on a Monte Carlo check of this relationship.
\begin{table}
\begin{center}
{\small
\begin{tabular}{c|c|c|c}
$F\cdot N^{3/4}$ & $N$ & $\langle r\rangle$ & fitted $c_2$ \\
\hline
\hline
 $13\tfrac{1}{3}$ & $10^4$ & $1663.54\pm 0.13$ & $0.7460\pm 0.0002$\\
\hline
$13\tfrac{1}{3}$ & $2\times 10^4$ & $2798.52\pm 0.19$ & $0.7454\pm 0.0002$\\
\hline
$13\tfrac{1}{3}$ & $4\times 10^4$ & $4705.25\pm 0.52$ & $0.7460\pm 0.0002$\\
\hline
$20$ & $10^4$ &  $1895.24\pm 0.16$ & $0.7471\pm 0.0002$\\
\hline
$20$  & $2\times 10^4$ & $3189.23\pm 0.32$ & $0.7457\pm 0.0003$\\
\hline
$20$ & $4\times 10^4$ & $5364.85\pm 0.64$ & $0.7453\pm 0.0003$\\
\hline
$30$ & $10^4$ & $2163.22\pm 0.28$ & $0.7484\pm 0.0003$\\
\hline
$30$ & $2\times 10^4$ & $3641.24\pm 0.58$ & $0.7463\pm 0.0003$\\
\hline
$30$ & $4\times 10^4$ & $6127.11\pm 1.03$ & $0.7451\pm 0.0004$
\end{tabular}
}
\caption{{\small Data from a Monte Carlo simulation of a self-avoiding walk with endpoints pulled apart with force $F$, so the weighting of a given walk is $e^{rF}$. For each value we used approximately 5 million walks, using a slightly modified version of Tom Kennedy's code \cite{kennedy2002faster}. The coefficient $c_2$ was fit assuming the form (\ref{app:form}). We are not sure what form to use in the extrapolation to infinite $N$, but ``by eye,'' we estimate $c_2 = 0.7445\pm 0.0007$.}}\label{tableOfData}
\end{center}
\end{table}

The coefficient $c_1$ is defined in (\ref{c1def}), and on the square lattice it has the value \cite{jensen2000size}
\be\label{c1valapp}
c_1= 3.39744\dots.
\ee
The coefficient $c_2$ is defined by the statement that the large $r$ asymptotics of the distribution for end-to-end distance of a self-avoiding walk of $N$ steps is 
\be\label{app:form}
P(r)\propto r^{1 + \frac{5}{8}}e^{-c_2 \frac{r^4}{N^3a^4}}.
\ee
In order to access the large $r$ region, we used the Metropolis algorithm to sample self-avoiding walks from the distribution $e^{Fr}P(r)$, where $F$ is a force that pulls the endpoints apart. In principle, one can fit the endpoint distribution with the force, $e^{F r}P(r)$, to binned Monte Carlo data. However, instead, we simply computed the expected value $\langle r\rangle$ in the ensemble with the force, and used this to fit the single parameter $c_2$ in the distribution (\ref{app:form}). 

From our data, we conclude that on the square lattice,
\be
c_2 = 0.7445\pm 0.0007.
\ee
See table \ref{tableOfData} for more details. The relationship (\ref{c2c1app}) together with the value (\ref{c1valapp}) predicts $c_2 = 0.745001\dots$. This is consistent with the value we found. The three digits of precision are not nearly as good as the tests of the RGJ formula in the small $y$ region \cite{richard2001scaling,jensen2004enumeration}, but this test still gives some additional confidence as it probes the opposite (large $y$) regime.

\section{CGHS model and flat space JT}\label{app: CGHS}
The CGHS model \cite{PhysRevD.44.314,PhysRevD.45.R1005,Strominger:1994tn} is a well studied two dimensional dilaton gravity model. Without matter fields, it has the action:
\be
I=-{1\over 2}\left[\int_{M} \sqrt{\tilde{g}}e^{-2\tilde\phi}\left( \tilde R+ 4(\tilde\nabla \tilde\phi)^2+4U\right)+2\int_{\partial M}\sqrt{\tilde{h}}e^{-\tilde\phi}\tilde K\right].
\ee 
Defining  $\lbrace g_{i j}=e^{-2\tilde \phi}\tilde g_{ij}, \phi=e^{-2\tilde\phi}\rbrace$ and using the Weyl transformation formulas
\be
\sqrt{\tilde{g}}\tilde R= \sqrt{g}(R+2\nabla^2 \tilde{\phi});~~~\sqrt{\tilde{h}}\tilde K=\sqrt{h}(K-\partial_{n}\tilde{\phi}),
\ee
one finds that this is equivalent to the action of flat space JT gravity \cite{dubovsky2017asymptotic}:
\be
I=-{1\over 2}\left[\int_{M} \sqrt{g}(\phi R+4U)+2\int _{\partial M}\sqrt{h}\phi_b K\right].
\ee 
$U$ is called the cosmological constant, and $\phi_b$ is the boundary value of the dilaton.  As in standard JT gravity, after integrating out the bulk dilaton filed, one gets an action for the shape of the boundary curve on a flat manifold:
\be
I=-2U A-2\pi \phi_b.
\ee
Note that in flat space JT gravity, the pressure term arises from the cosmological constant, not the boundary value of the dilaton. But in any case, we get the same area action plus a constant $2\pi \phi_b$, and the self avoiding walk analysis does not change. So, using the RGJ formula, the exact density of states for CGHS model without matter is given by:
\be\label{for: Airry}
\density(E)={1\over 4\pi^4}{(2U)^{1\over 3}e^{2\pi \phi_b}\over \text{Ai}^2(-E (2U)^{-{2\over 3}})+\text{Bi}^2(-E (2U)^{-{2\over 3}})}.
\ee   

We would like to understand how this can be consistent with the statement that the CGHS model has a fixed temperature, and thus an entropy that is linear in the energy. The main point is that this will be consistent with (\ref{for: Airry}) in an asymptotic limit. Let's first discuss this limit for classical flat space JT gravity, with solution
\be
\mathrm{d}s^2 = \mathrm{d}r^2 + r^2\mathrm{d}\theta^2, \hspace{20pt}
\phi=\phi_h+U r^2.
\ee
To take an asymptotic limit, the boundary length $L = 2\pi r_b$ goes to infinity and the boundary value of the dilaton also goes to infinity, with $\phi_h$ fixed. The classical solution implies a relationship between $L$ and $\phi_b$ in this limit:
\be\label{rel}
\phi_b\sim U r_b^2=U \left(\frac{L}{2\pi}\right)^2.
\ee
To define the inverse temperature, we are allowed to rescale the length $L$ by some factor of $\phi_b$. To get a finite answer, we should divide by $\sqrt{\phi_b}$. Including also a factor of $\sqrt{U}$ for convenience, we set $\beta = L/\sqrt{\phi_b U}$. Then (\ref{rel}) implies $\beta = {U\over 2\pi}$ for any $\phi_h$  \cite{Maldacena:2019cbz}.

Let's now discuss this for the quantum theory based on self-avoiding walks. Using the large-pressure relation (\ref{eom}), we find
\be\label{asym}
{U\upbeta^3\over 3}=\pi r_b^2\ell^2\sim {\pi\phi_b\over U}.
\ee
We can define a rescaled inverse temperature $\beta$ that is finite in the asymptotic limit as $\beta = \upbeta / \phi_b^{1/3}$. Then (\ref{asym}) implies
\be\label{for:tildebeta}
\beta={(3\pi)^{1/3}\over U^{2/3}},
\ee
so again the temperature has a fixed value. 

In terms of the energy, the asymptotic limit is a limit where we take the energy large and negative, and then study relatively small deviations from this value. In this ``zoomed-in'' limit, the logarithm of (\ref{for: Airry}) is approximately a linear function of the energy, consistent with the expectations. However, it is clear that the asymptotic limit misses many details of the full function (\ref{for: Airry}).

\section{Large unpressurized loops}\label{app:large unpressurized}
The regimes we analyzed above don't cover the case where the pressure is much smaller than the AdS scale, but the loop is much larger than the AdS scale. In this regime we expect $Z\propto \upbeta^{-3/2}$.

One piece of evidence for this is that the answer should not be very sensitive to the self-avoiding constraint. This is because on scales larger than the AdS scale, ordinary random walks in hyperbolic space tend not to self-intersect. And for the ordinary random walk, the results of \cite{Kitaev:2018wpr,Yang:2018gdb} imply that the answer is proportional to $\upbeta^{-3/2}$.

This power-law is supported by exact enumeration data of self-avoiding loops on a coarse lattice regularization of hyperbolic space \cite{swierczak1996self}, where the power $-3/2$ is confirmed to within percent-level precision. This power is also the right one to match smoothly with the answers in the adjacent Schwarzian limit.

{\footnotesize
\bibliography{references}
}

\bibliographystyle{utphys}

\end{document}